\documentclass[journal=jctcce,manuscript=article]{achemso}

\usepackage{chemformula} 
\usepackage[T1]{fontenc} 
\usepackage{booktabs}
\usepackage{amsmath}

\let\oldAA\AA
\renewcommand{\AA}{\text{\normalfont\oldAA}}

\author{Jonah Marks}
\author{Joseph Gomes}
\email{joe-gomes@uiowa.edu}
\affiliation[Iowa]
{Department of Chemical and Biochemical Engineering \\University of Iowa, Iowa City, United States}

\title[fsm]
  {Incorporation of Internal Coordinates Interpolation into the Freezing String Method}

\abbreviations{MEP,TS,FSM,LST,RIC}
\keywords{Chemical calculations, Chemical reactions, Optimization, Transition states}

\begin{document}

\begin{tocentry}

\includegraphics[width=8.5cm]{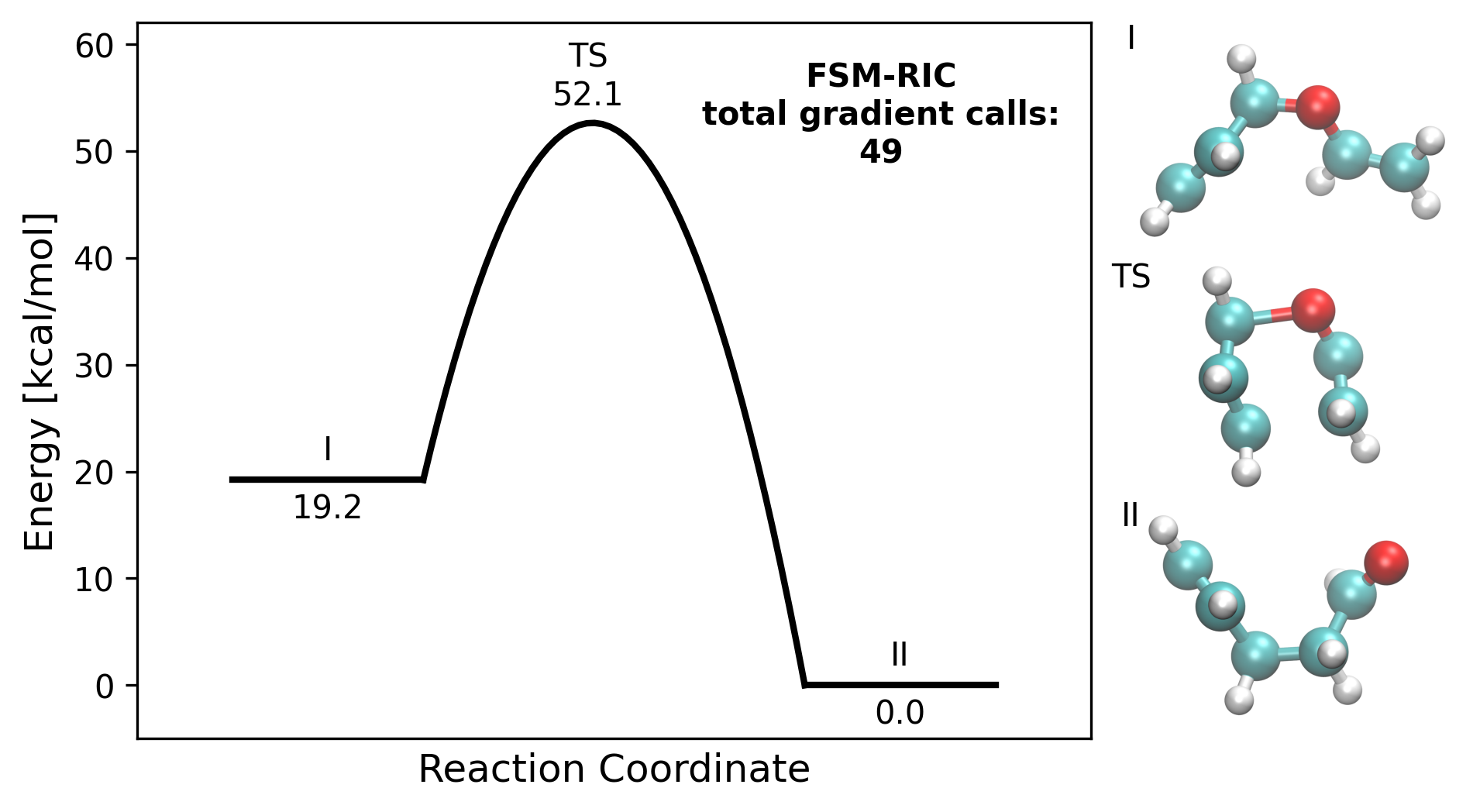}\\

Reaction coordinate diagram of the CH$_2$CHCH$_2$CH$_2$CHO Claisen rearrangement reaction. The transition state search starting from the FSM-RIC method required fewer than 50 DFT gradient evaluations in total.

\end{tocentry}

\begin{abstract}
We present an improved method for determining guess structures for transition state searches by incorporating internal coordinates interpolation into the freezing string method (FSM). We test our method on over 40 reactions across 3 benchmark datasets covering a diverse set of chemical reactions. Our results show that incorporation of internal coordinates interpolation improves the reliability of the FSM, enabling larger interpolation step sizes and fewer optimization steps per cycle, which together yield nearly a 50\% reduction in computational cost while maintaining a 100\% success rate on benchmark chemical reaction test cases, including systems where previous attempts based on linear synchronous transit interpolation have failed. We provide an open-source Python implementation of the FSM, in addition to the reactant, product, and transition state structures of all reactions studied.
\end{abstract}

\section{Introduction}
The determination of reaction mechanisms is an important step for chemists in the prediction of thermodynamics and kinetic properties of chemical reactions. In computational chemistry, the reaction mechanism is typically represented as a pathway on the Born-Oppenheimer potential energy surface (PES) of the system of interest given by a suitable potential energy function of the nuclear coordinates. The equilibrium states correspond to stationary points on the PES with zero first order derivatives (gradient) in all directions and all positive eigenvalues of the second order derivative (Hessian) matrix. The equilibrium states can be identified by energy minimization given a suitable initial guess structure. There exist many pathways by which equilibrium states may interconvert; however, only a small subset of these pathways are thermally accessible and relevant to thermodynamic and kinetic analysis of reaction mechanisms. The minimum energy path (MEP) is the route that needs the least amount of potential energy for the system to undergo the transition. The transition states (TSs), based on transition state theory, are first order saddle points with zero gradient and only one negative eigenvalue of the Hessian matrix.  The MEP connecting two equilibrium states must go through one or more TSs and serves as a representative reaction path.

The first step in identifying the MEP is typically locating the minimum energy TS connecting the equilibrium states of interest. Determination of TS geometries is a computationally expensive task that frequently requires significant human intervention. Efforts have been made to develop automated processes for finding first order saddle points on PESs while attempting to minimize computational cost. Many modern algorithms take a chain-of-states\cite{pratt1986statistical, elber1987method} approach to this problem. Geometric interpolation between the reactant and product geometries is performed, producing a string of intermediate structures, or nodes, distributed along the reaction pathway. The resulting structures are optimized using ab-initio methods to approximate the MEP of the reaction. The highest energy structure in the chain-of-states is taken as a guess of the TS geometry. A local surface walking algorithm refines the guess to exact TS geometry. \cite{cerjan1981finding,schlegel1982optimization,simons1983walking, baker1986algorithm,wales1992basins,wales1993locating,banerjee1985search, ayala1997combined, heyden2005efficient} These algorithms require the use of local gradient and, in some cases, Hessian information to locate saddle points, and their success is heavily dependent on the initial guess being located within the desired basin of attraction.  

Many such chain-of-states methods that provide accurate TS geometry guesses from reactant and product structures have been previously reported. Details on how intermediate structures are selected and optimized differs between methods, and consequently influences the computational cost and success of the algorithm. The nudged elastic band (NEB) method creates an initial chain-of-states by interpolation between reactant and product, and every intermediate structure along the string is iteratively optimized to lie along the reaction pathway, together with an additional penalty term to maintain an even distribution of structures along the chain-of-states. \cite{mills1994quantum, henkelman2000improved, henkelman2000climbing, maragakis2002adaptive, chu2003super, trygubenko2004doubly, kolsbjerg2016automated, ruttinger2022protocol} The growing string method (GSM) in most use cases incurs fewer gradient calculations than the NEB by developing a better initial chain-of-states, or string.\cite{peters2004growing, quapp2005growing, goodrow2008development, goodrow2009transition, goodrow2010strategy, behn2011incorporating, zimmerman2013growing, zimmerman2013reliable, zimmerman2015single, jafari2017reliable} The GSM creates two strings; one starts at the reactant configuration and the other starts at the product configuration. At each iteration of the growing process, an intermediate structure, or node, is added to each string frontier in the direction of the opposing string, after which the entire string is optimized. This growth process is repeated until the strings meet, providing a better initial chain-of-states, and followed by additional optimization of the unified string. In addition to providing a better initial chain-of-states, this avoids simulation of non-physical structures in the interior nodes which may result from the initial interpolation. 

The freezing string method (FSM) further reduces the number of gradient calls but at the cost of identifying the true MEP. \cite{behn2011efficient, mallikarjun2012automated, suleimanov2015automated} Two strings are grown from the product and reactant like the GSM. Once a frontier node is placed, optimization is performed to step the node closer to the reaction pathway, after which it is “frozen”, i.e., it will not move for the remainder of the calculation. New frontier nodes are added to both strings, and the process is repeated. Once the two strings unite, the highest energy structure is taken to be the TS guess without optimization of the unified string. In practice, the resulting pathway can deviate significantly from the MEP but often produces guess structures suitable for further refinement.

The FSM greatly improves the computational efficiency of TS guess finding compared to previously described algorithms, often resulting in an order of magnitude reduction in the number of electronic structure calculations required to determine the true TS structure. Despite these improvements, the FSM is not a foolproof algorithm. The efficiency of the calculation and ultimately its success heavily depends on the interpolation algorithm used to generate initial guess structures at the frontier of the growing strings. There exist known issues with commonly applied Cartesian coordinate and linear synchronous transit (LST) interpolation techniques\cite{halgren1977synchronous}, which can produce high energy or otherwise aphysical molecular geometry structures far from the true MEP which are then subjected to electronic structure calculation and geometry optimization.\cite{zimmerman2013growing, smidstrup2014improved} Due to the relatively few number of optimization steps used in the FSM, the incorporation of these geometries as anchor points into future interpolation steps poisons the calculation, resulting in a failed search and wasted computational effort.

In this work, we demonstrate an improved method for initiating TS searches by incorporating internal coordinates (ICs) interpolation into the FSM. Additionally, we incorporate the L-BFGS-B with explicit line search for step size determination which improves reaction pathway optimization steps. We test our method on over 40 reactions across three benchmark datasets covering a broad set of chemical reactions. Our results show that, using previously studied LST interpolation, the incorporation of the L-BFGS-B optimization reduces the computational effort required in previous studies, even considering the additional computational cost incurred due to step size determination by line search. Incorporation of ICs interpolation further improves the computational efficiency of the FSM and successfully locates high-quality TS guess structures where previous attempts based on LST interpolation have failed. We provide an open-source Python implementation of the FSM, in addition to the reactant, product, TS structures of all reactions computed at the $\omega$B97X-V/def2-TZVP level of theory. We anticipate that these resources will be of broad interest for researchers in computational chemistry studying problems where the fast and reliable location of TSs is important, or those developing algorithms who wish to evaluate their methods on a broad set of realistic chemical problems.

\section{Methods}
\subsection{Overview of the Freezing String Method}
\begin{figure}[t]
\centering
\includegraphics[width=9cm]{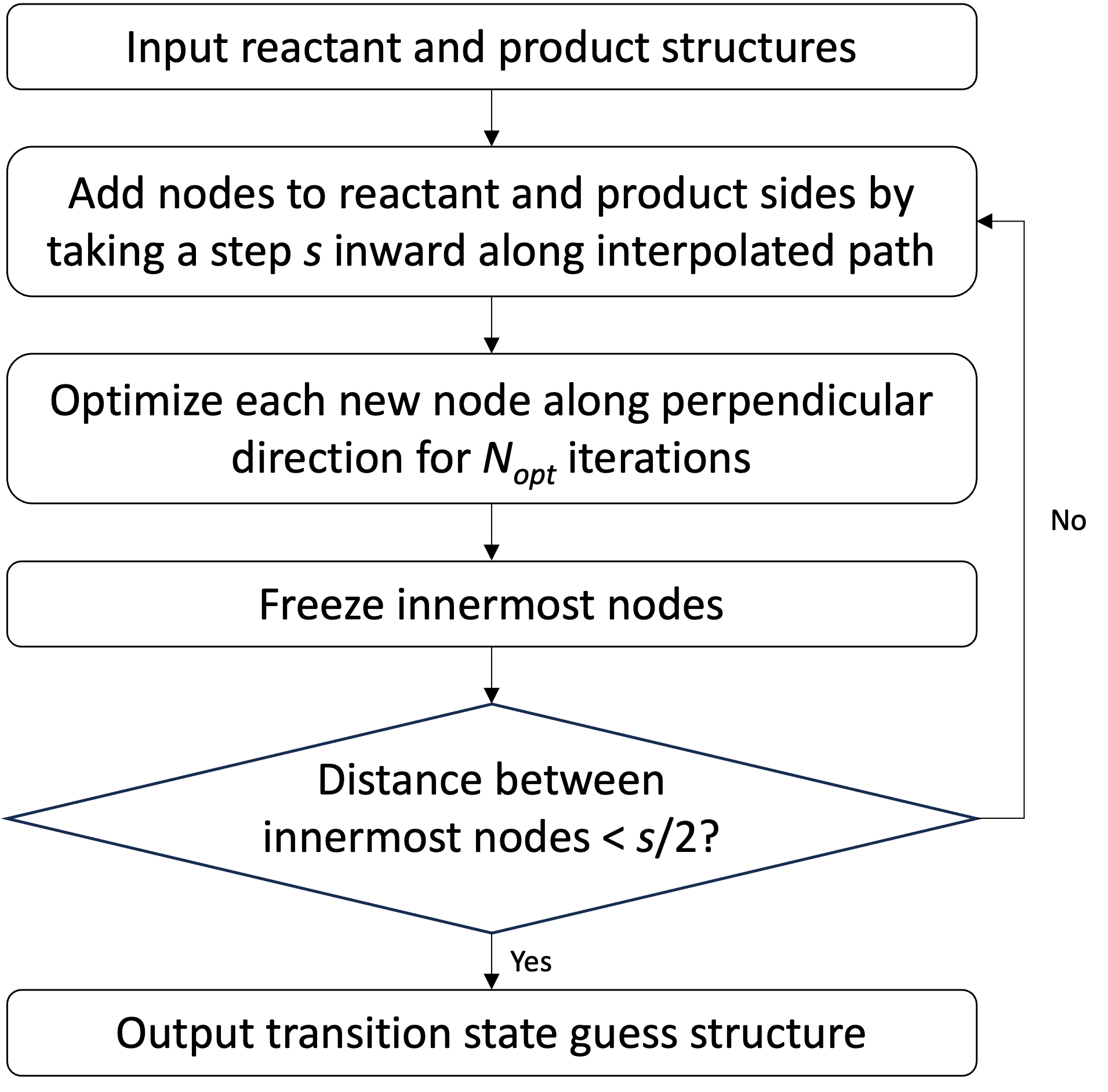}
\caption{
Algorithm flowchart describing the freezing string method. $s$ is the interpolation step size and $N_\mathrm{opt}$ is the maximum number of optimization steps per interpolation step.
}
\label{fig:flowchart}
\end{figure}

A flowchart of the FSM algorithm is shown in Figure \ref{fig:flowchart}. The goal of the FSM is to produce an approximate reaction pathway connecting two given reactant and product structures. The approximate reaction pathway is represented as a chain-of-states consisting of nodes along a parameterized string. The interior nodes on the string represent intermediate geometries along the approximate reaction pathway. The method evolves the string by alternately adding nodes to the reactant and product sides of a growing string. The new reactant and product side structures are generated by taking a step inward along an interpolated path between frontier nodes. After interpolation, the structures undergo geometry optimization in the direction perpendicular to the approximate reaction pathway. Optimization of these new frontier structures is performed, and then the geometries are frozen. These steps are repeated until the reactant and product sides of the growing string meet. The highest energy node along the pathway is chosen as the TS guess structure that is refined to the true TS geometry using a local surface walking optimization algorithm.

Prior to beginning the FSM calculation, the reactant and product structures are first aligned using the Kabsch algorithm to minimize rigid-body rotation and translation. During the first evolution step, the frontier nodes are the user-defined reactant and product structures, while subsequent steps use the frozen nodes from the previous iteration as anchor points for interpolation. The interpolation step size $s$ is fixed during the calculation, and in this work, we determine $s$ by dividing the arc length of an interpolated path connecting the initial reactant and product structures by a user-defined nominal number of nodes, $N_{\mathrm{nodes}}$.The interpolation step is then performed by adding nodes at this distance, $s$, from the reactant and product frontier nodes. The interpolation can be performed using many different coordinate systems. We consider both LST interpolation and linear interpolation in redundant internal coordinates. Tangent directions at the new reactant and product frontier nodes are determined from a cubic spline fitted through the coordinates of the entire interpolated pathway, starting at the previously frozen nodes, as a function of path arc length.

During optimization, the energy is minimized by assumption of the local quadratic approximation:

\begin{equation}
E(\boldsymbol{x}) = E(\boldsymbol{x_0}) + (\boldsymbol{x} - \boldsymbol{x_0})^T \boldsymbol{g}^{\perp} 
+\frac{1}{2}(\boldsymbol{x} - \boldsymbol{x_0})^T \boldsymbol{H} (\boldsymbol{x} - \boldsymbol{x_0})
\label{eqn:energy}
\end{equation}

Here, $\boldsymbol{x_0}$ is the current geometry, $\boldsymbol{x}$ is a proposed geometry, $\boldsymbol{g}^{\perp}$ is the perpendicular gradient, and $\boldsymbol{H}$ is the approximate Hessian in the space perpendicular to the approximate reaction pathway. The perpendicular gradient is given by $\boldsymbol{g}^{\perp} = (\boldsymbol{I}-\boldsymbol{\hat{t}}\boldsymbol{\hat{t}}^T) \boldsymbol{g}$ where $\boldsymbol{\hat{t}}$ is the normalized tangent vector determined during the interpolation step. 

The optimization of equation \ref{eqn:energy} is performed using the L-BFGS-B algorithm\cite{liu1989limited,byrd1995limited} as implemented in the SciPy Python library.\cite{virtanen2020scipy} The L-BFGS-B algorithm retains 10 vectors for the Hessian approximation. Optimizations are performed in Cartesian coordinates in this work. We place bounds on each Cartesian coordinate such that no coordinate is displaced by more than 0.3 $\AA$  during a single optimization step. Each optimization step is begun by performing a backtracking line search, where at most $N_{\mathrm{ls}}$ (often one) energy and gradient calculations are performed to determine an appropriate step size. The optimization proceeds until a specified convergence criteria is satisfied, or a maximum number of step $N_{\mathrm{opt}}$ has been performed. 

\subsection{Redundant internal coordinates} Cartesian coordinates $\boldsymbol{x}= (x_1, x_2, \dots, x_{3N})$ are used to specify the positions of $N$ atoms in a molecule and are a necessary input for electronic structure calculations. Internal coordinates (ICs) $\boldsymbol{q} = (q_1(\boldsymbol{x}), q_2(\boldsymbol{x}), \dots, q_n(\boldsymbol{x}))$ are functions of the Cartesian coordinates that better describe the collective motions of atoms. There exists several choices for constructing coordinate systems when performing interpolation between two molecular geometries. The definition of the redundant internal coordinates used in this work have been reported previously \cite{pulay1992geometry,peng1996using,Wang2016}. Redundant internal coordinate sets are constructed separately for reactant and product molecules, and the union of these two sets gives the full set of redundant coordinates. Bonds, angles, linear angles, torsion angles, and out-of-plane angles are assigned based on a procedure outlined in \citet{bakken2002efficient}.

The full set of coordinates is further pruned using the following set of heuristics to ensure each internal coordinate is well-defined for both reactant and product molecules. If a given bond angle is nearly linear in either the reactant or product geometry ($\angle_{\mathrm{ABC}}>175^o$), the angle is removed and replaced by two orthogonal linear angle bending coordinates which ensure that the linear structure is stabilized. A linear angle bending coordinate is pruned if the molecule undergoes transformation such that a linear bending angle $\angle_{\mathrm{ABC}}>45^o$ is present in reactant or product geometries. Torsion coordinate angles $\angle_{\mathrm{ABCD}}>175^o$ in either the reactant or product structure are removed and a bonding coordinate between atoms A and D is added. Torsion coordinates containing atoms forming bending angles $\angle_{\mathrm{ABC}}>175^o$ or $\angle_{\mathrm{BCD}}>175^o$ in either the reactant or product structure are removed. Out-of-plane bending angles $\angle_{\mathrm{ABCD}}>175^o$ in either the reactant or product structure are removed, as well as any out-of-plane angles containing broken bonding centers. Finally, if the number of atoms $N$ is greater than 3 and no torsions remain after pruning, all unique permutations of atoms A, B, C, and D are tried until a valid torsion angle $\angle_{\mathrm{ABCD}}<175^o$ satisfying previous criteria is identified. If still no torsions are found, the set of internal coordinates is replaced by the set of all unique atom-atom distances.

\subsection{ICs interpolation}

Given a set of $n$ primitive redundant coordinates $\boldsymbol{q}$, a set of structures $\boldsymbol{q}^i(f)$ are produced by linear interpolation

\begin{equation}
    \boldsymbol{q}^i(f) = (1-f)\boldsymbol{q}^R + f \boldsymbol{q}^P
\end{equation}

where $f$ is the interpolation parameter and $\boldsymbol{q}^R$ and $\boldsymbol{q}^P$ represent the reactant and product internal coordinate values. The displacements $\boldsymbol{\Delta q}$ in $\boldsymbol{q}$ are related by the well-known $B$ matrix\cite{wilson1980molecular} $\boldsymbol{\Delta q} = \boldsymbol{B} \boldsymbol{\Delta x}$ valid for small Cartesian displacements $\boldsymbol{\Delta x}$. We form the full $\boldsymbol{B}^{\mathrm{prim}}$ matrix for all $n$ primitive internals. We identify and remove linearly dependent rows of $B$ by forming and diagonalizing the matrix $\boldsymbol{G}=\boldsymbol{B}^{\mathrm{prim}}(\boldsymbol{B}^{\mathrm{prim}})^T$. The diagonalization of $\boldsymbol{G}$ yields two sets of eigenvectors spanning the nonredundant and redundant subspace of our original space of primitive internals $\boldsymbol{q}$. The eigenvalue equation of $\boldsymbol{G}$ is

\begin{equation}
    \boldsymbol{G}(\boldsymbol{U}\boldsymbol{R}) = (\boldsymbol{U}\boldsymbol{R})\begin{pmatrix}
        \boldsymbol{\Lambda}&\boldsymbol{0}\\
        \boldsymbol{0}&\boldsymbol{0}
    \end{pmatrix}
\end{equation}

where $\boldsymbol{U}$ is the set of nonredundant eigenvectors of $\boldsymbol{G}$ and $\boldsymbol{R}$ is the corresponding redundant set. The full $\boldsymbol{B}^{\mathrm{prim}}$ matrix is transformed to our active, delocalized internal coordinate set by

\begin{equation}
    \boldsymbol{B} = \boldsymbol{U}^T\boldsymbol{B}^{\mathrm{prim}}.
\end{equation}

The inverse $B$ matrix is constructed

\begin{equation}
    (\boldsymbol{B}^T)^{-1} = (\boldsymbol{B}\boldsymbol{B}^T)^{-1}\boldsymbol{B}
\end{equation}

and used to convert displacement in internal coordinates $\boldsymbol{\Delta q}^{\mathrm{dlc}} = \boldsymbol{U}^T \boldsymbol{\Delta q}$ to Cartesian

\begin{equation}
    \boldsymbol{\Delta x} = (\boldsymbol{B}^T)^{-1} \boldsymbol{\Delta q}^{\mathrm{dlc}}.
\end{equation}

The final step is to transform the geometries along the interpolated path from internal coordinates back to Cartesians. Given a target geometry $\boldsymbol{q}^i$ in internal coordinates, this is done iteratively using the formula

\begin{equation}
    \boldsymbol{x}_{k+1} = \boldsymbol{x}_k + (\boldsymbol{B}(\boldsymbol{x}_k)^T)^{-1}[\boldsymbol{q}^{\mathrm{dlc},i}-\boldsymbol{q}^{\mathrm{dlc}}(\boldsymbol{x}_k)].
\end{equation}

The iteration is terminated when the Cartesian coordinates generated on the ($k+1$)th iteration $\boldsymbol{x}_{k+1}$ are identical to those on the $k$th iteration $\boldsymbol{x}_k$ within a tolerance $10^{-7}$ $\AA$\cite{bakken2002efficient}. The transformation of all $\boldsymbol{q}^i(f)$ results in a chain-of-states $\boldsymbol{x}^c(f)$ connecting the current frontier reactant and product nodes that is used during the interpolation step of the FSM.

\subsection{LST interpolation} LST is another chemically realistic interpolation method that produces a chain-of-states pathway.\cite{halgren1977synchronous}  A set of structures $\boldsymbol{r}^i(f)$ are produced by linear interpolation in a set of internuclear distances $\boldsymbol{r}^i=\{r_{ab}; a>b=1, 2, \dots, N\}$

\begin{equation}
    \label{eqn:r_interp}
    \boldsymbol{r}^i(f) = (1-f)\boldsymbol{r}^R + f \boldsymbol{r}^P
\end{equation}

where $f$ is the interpolation parameter and $\boldsymbol{r}^R$ and $\boldsymbol{r}^P$ represent the reactant and product internuclear distances. A set of structures $\boldsymbol{x}^i(f)$ are produced by linear interpolation in Cartesian coordinates

\begin{equation}
    \boldsymbol{x}^i(f) = (1-f)\boldsymbol{x}^R + f \boldsymbol{x}^P
\end{equation}

The final interpolated pathway is determined by minimizing the objective $S^{\mathrm{LST}}$ by the method of least-squares at each interpolation point $f$

\begin{equation}
    S^{\mathrm{LST}} = \sum_{a>b}^N \frac{(r_{ab}^i-r_{ab}^c)^2}{(r_{ab}^i)^4} + w \sum_{j=1}^{3N} (x_{j}^i-x_{j}^c)^2
\label{eqn:lst}
\end{equation}
where $w$ is a weighting factor (nominally $10^{-6}$) and the superscripts $i$ and $c$ denote interpolated and calculated values, respectively. The variables $\boldsymbol{x}^c$ are those optimized during the minimization of $S^{\mathrm{LST}}$, and the variables $\boldsymbol{r}^c$ are derived from $\boldsymbol{x}^c$. The denominator in the first term ensures that important distances between bonded atoms are preserved in the final interpolated geometries. The second term is weighted by a factor $w$, and produces forces to prevent rigid translation or rotation such that the interpolated structures align with the end structures. The weighting factor $w$ additionally weights within the objective function the relative importance of linear interpolation in internuclear distances (first term) and linear interpolation in Cartesian coordinates (second term). The optimization of equation \ref{eqn:lst} is performed using the L-BFGS-B algorithm\cite{byrd1995limited} as implemented in the SciPy Python library. The optimization of $S$ for all $f$ results in a chain-of-states $\boldsymbol{x}^c(f)$ connecting the current frontier reactant and product nodes that is used during the interpolation step of the FSM.

\subsection{Computational details} Electronic structure calculations are performed to produce the optimized geometries of the reactant and product for each reaction studied, compute the quantum mechanical gradients required by the chain-of-states algorithms, as well as to refine the geometry of the TS guess structures. All electronic structure calculations were performed using the range-separated, hybrid generalized gradient approximation with non-local correlation exchange-correlation function $\omega$B97X-V\cite{mardirossian2014omegab97x} using the triple-$\zeta$, polarized valence def2-TZVP basis set.\cite{weigend2005balanced} During geometry optimization, energies were converged to 10$^{-6}$ Ha (Hartree) and the maximum norm of the Cartesian gradient was converged to 10$^{-3}$ Ha bohr$^{-1}$. Geometry optimization was terminated if the convergence criteria were not met within 250 optimization steps. All reported energy values are electronic energy without thermal or zero-point correction. The eigenvector-following local TS searches were performed using the partitioned rational function optimization method initialized with analytical Hessian calculation at the guess geometry.\cite{baker1986algorithm} Frequency analysis was performed to confirm the nature of each stationary point: there must be zero imaginary frequencies for PES minima and exactly one imaginary frequency for PES TSs. Intrinsic reaction coordinate (IRC) pathway calculations initiated at the TS were performed with the predictor-corrector algorithm of \citet{schmidt1985intrinsic} to further characterize TS geometries. The FSM software is written in Python and performs file-based data exchange with an electronic structure software package to obtain the quantum mechanical gradients used in the method. All QM calculations were performed using a release version of the Q-Chem 6.0 software package.\cite{epifanovsky2021software}

\section{Results and Discussion}

We measure benchmark performance by tracking the number of quantum mechanical gradient calculations performed as part of the FSM and TS search procedure, as well as the total number of gradient calculations. We report separately the total number of gradients computed during FSM calculation, which measures the efficiency at which a TS guess is obtained, and the total number of gradients computed during the TS calculation, which serves as an indicator for TS guess quality. We compare FSM calculations with either redundant internal coordinates (FSM-RIC) or linear synchronous transit (FSM-LST) interpolation methods. We perform FSM calculations with nominally 18 nodes ($N_{\mathrm{nodes}}$) along the approximate reaction pathway, two steps per optimization cycle ($N_{\mathrm{opt}}$), and at most three line search steps per optimization cycle ($N_{\mathrm{ls}}$) unless otherwise stated. We consider this set of parameters to be a conservative baseline where both the FSM-RIC and FSM-LST should perform well.

\subsection{Sharada test set}

\begin{table*}[ht]
\centering
\resizebox{\textwidth}{!}{\begin{tabular}{|l|l|r|r|r|r|r|r|}
\toprule
ID & Reaction & \multicolumn{1}{|p{2cm}|}{\centering Gradients \\ (FSM-RIC)} & \multicolumn{1}{|p{2cm}|}{\centering Gradients \\ (TS-RIC)} & \multicolumn{1}{|p{2cm}|}{\centering Gradients \\ (TOTAL-RIC)} & \multicolumn{1}{|p{2cm}|}{\centering Gradients \\ (FSM-LST)} & \multicolumn{1}{|p{2cm}|}{\centering Gradients \\ (TS-LST)} & \multicolumn{1}{|p{2cm}|}{\centering Gradients \\ (TOTAL-LST)}\\
\midrule
1 & H$_2$CO $\rightarrow$ H$_2$ + CO & \textbf{58} & \textbf{13} & \textbf{71} & 62 & 11 & 73 \\
2 & SiH$_2$ + H$_2$ $\rightarrow$ SiH$_4$ & 55 & 5 & 60 & \textbf{53} & \textbf{4} & \textbf{57} \\
3 & acetaldehyde Keto-enol tautomerism & 63 & 3 & 66 & \textbf{61} & \textbf{4} & \textbf{65} \\
4 & CH$_3$CH$_3$ $\rightarrow$ CH$_2$CH$_2$ + H$_2$ & 58 & 24 & 82 & \textbf{52} & \textbf{21} & \textbf{73} \\
5 & bicyclo[1.1.0]butane $\rightarrow$ $\textit{trans}$-butandiene & 61 & 47 & 108 & \textbf{58} & \textbf{25} & \textbf{83} \\
6 & parent Diels-Alder cycloaddition reaction & \textbf{62} & \textbf{15} & \textbf{77} & 53 & 33 & 86 \\
7 & $\textit{cis}$,$\textit{cis}$-2,4-hexadiene $\rightarrow$ 3,4-dimethylcyclobutene  & \textbf{60} & \textbf{18} & \textbf{78} & 79 & 25 & 104 \\
8 & alanine dipeptide rearrangment & 61 & 64 & 125 & \textbf{45} & \textbf{45} & \textbf{90} \\
9 & silyl ketene acetal $\rightarrow$ silyl ester Ireland-Claisen rearrangement & \textbf{60} & \textbf{82} & \textbf{142} & 60 & 88 & 148 \\
\bottomrule
~ & Average & 60 & 30 & 90 & 58 & 28 & 87 \\
\bottomrule
\end{tabular}}
\caption{Comparison of performance of the FSM-RIC and FSM-LST for TS guess structure generation and TS search on the Sharada benchmark set. Performance is measured based on the total number of gradient evaluations required to achieve FSM and TS search convergence. The method (FSM-RIC or FSM-LST) requiring the fewest gradient evaluations is highlighted in bold text.} 
\label{table:sharada}
\end{table*} 

A test suite consisting of nine reactions curated by \citet{mallikarjun2012automated} is chosen as the first benchmark. The Sharada test set contains a broad set of bond formation, dissociation, ring-opening, and isomerization reactions. The description of the nine reactions in the Sharada test set is presented in Table \ref{table:sharada}. This choice of benchmark suite is significant due to its previous use benchmarking previous implementations of the FSM\cite{mallikarjun2012automated,behn2011incorporating}. For each reaction, we highlight in bold the method (FSM-RIC or FSM-LST) that locates the TS geometry in fewest calculations.

The comparison between the required gradient calls for FSM and TS calculations using RIC or LST interpolation on the Sharada test set is presented in Table \ref{table:sharada}. Both FSM-RIC and FSM-LST calculations successfully locate the exact benchmark TS structure without user intervention when performed with the conservative baseline parameters. The FSM-RIC and FSM-LST perform similarly, on average, producing a TS guess structure after 60 and 58 gradient evaluations, respectively. The resulting TS guess structures are of similar quality, as indicated by the average number of gradient evaluations required for TS structure refinement. The FSM-RIC TS guess structure requires on average 30 gradient evaluations for subsequent optimization, and the FSM-LST TS guess structures require on average 28 gradient evaluations for optimization.

Though the choice of exchange-correlation method and basis set in the original study by \citet{mallikarjun2012automated} precludes the direct comparison to the current work, we can qualitatively compare their reported gradient evaluation counts to ours. The average number of gradient calls required for the FSM-LST in its original implementation is 53, noticeably fewer than required by our implementation of FSM-LST, when performed with similar FSM parameter settings. The two implementations differ notably in how each method chooses the step size for node-level optimization. Whereas the original work uses the BFGS algorithm and a heuristic for step size determination based on available local PES information, our method uses the L-BFGS-B algorithm together with a step size determined by explicit line search. There is a small increase in the number of gradient calls required during the FSM calculation due to the explicit line search requiring additional electronic structure calculations at each optimization step. The average number of gradient calls required for subsequent TS optimization in the previous FSM implementation is 64, whereas our method requires on average 28 gradient calls for TS optimization. We can reasonably conclude that, although our optimization method requires more gradient calls due to explicit line search, our TS guesses are of higher quality and converge to the correct TS structure in overall fewer gradient calls on average.

\subsection{Birkholz test set}

\begin{table*}[ht]
\centering
\resizebox{\textwidth}{!}{\begin{tabular}{|l|l|r|r|r|r|r|r|r|r|}
\toprule
ID & Reaction & \multicolumn{1}{|p{2cm}|}{\centering Gradients \\ (FSM-RIC)} & \multicolumn{1}{|p{2cm}|}{\centering Gradients \\ (TS-RIC)} & \multicolumn{1}{|p{2cm}|}{\centering Gradients \\ (TOTAL-RIC)} & \multicolumn{1}{|p{2cm}|}{\centering Gradients \\ (FSM-LST)} & \multicolumn{1}{|p{2cm}|}{\centering Gradients \\ (TS-LST)} & \multicolumn{1}{|p{2cm}|}{\centering Gradients \\ (TOTAL-LST)}\\
\midrule
1 & C$_2$H$_4$ + N$_2$O $\rightarrow$ C$_2$N$_2$O & \textbf{69} & \textbf{7} & \textbf{76} & 61 & 48 & 109 \\
2 & 1,3-pentadiene hydrogen transfer & \textbf{63} & \textbf{5} & \textbf{68} & 68 & 6 & 74 \\
3 & HCN $\rightarrow$ HNC & \textbf{78} & \textbf{3} & \textbf{81} & \textbf{72} & \textbf{9} & \textbf{81} \\
4 & 1,4-hexadiene Cope rearrangement & 56 & 27 & 83 & \textbf{60} & \textbf{12} & \textbf{72} \\
5 & 1,3-cyclopentadiene hydrogen shift & \textbf{62} & \textbf{5} & \textbf{67} & 64 & 7 & 71 \\
6 & 1,3-butadiene cyclization & \textbf{60} & \textbf{6} & \textbf{66} & 66 & 8 & 74 \\
7 & Diels-Alder endo addition of cyclopentadiene to cyclopentadiene & 62 & 50 & 112 & \textbf{57} & \textbf{52} & \textbf{109} \\
8 & Diels-Alder addition of cyclopentadiene and ethylene & \textbf{62} & \textbf{11} & \textbf{73} & 63 & 11 & 74 \\
9 & difluorocarbene addition to ethylene & 56 & 24 & 80 & \textbf{50} & \textbf{12} & \textbf{62} \\
10 & ene reaction of ethylene and propene & \textbf{49} & \textbf{42} & \textbf{91} & 51 & 47 & 98 \\
11 & Grignard addition of phenyl magnesium bromide to benzophenone & \textbf{53} & \textbf{38} & \textbf{91} & 59 & 49 & 108 \\
12 & H$_2$CO $\rightarrow$ H$_2$ + CO & \textbf{58} & \textbf{13} & \textbf{71} & 62 & 11 & 73 \\
13 & CH$_3$CH$_2$F $\rightarrow$ CH$_2$CH$_2$ + HF & \textbf{52} & \textbf{9} & \textbf{61} & 67 & 9 & 76 \\
14 & water assisted hydrolysis of ethyl acetate & \textbf{42} & \textbf{60} & \textbf{102} & 56 & 86 & 142 \\
15 & H$_2$ + H$_2$CO $\rightarrow$ CH$_3$OH & \textbf{52} & \textbf{8} & \textbf{60} & 61 & 17 & 78 \\
16 & 2-methyl-3-phenyloxirane ring opening & 58 & 48 & 106 & \textbf{59} & \textbf{36} & \textbf{95} \\
17 & CH$_2$CHCH$_2$CH$_2$CHO Claisen rearrangement & \textbf{57} & \textbf{31} & \textbf{88} & 61 & 30 & 91 \\
18 & SiH$_2$ + H$_2$ $\rightarrow$ SiH$_4$ & 55 & 5 & 60 & \textbf{53} & \textbf{4} & \textbf{57} \\
19 & Cl$^-$ + CH$_3$F $\rightarrow$  CH$_3$Cl + F$^-$& \textbf{56} & \textbf{4} & \textbf{60} & 60 & 3 & 63 \\
20 & sulfur dioxide addition to butadiene & \textbf{63} & \textbf{21} & \textbf{84} & 65 & 22 & 87 \\
\bottomrule
~ & Average & 58 & 21 & 79 & 61 & 24 & 85 \\
\bottomrule
\end{tabular}}
\caption{Comparison of performance of the FSM-RIC and FSM-LST for TS guess structure generation and TS search on the Birkholz benchmark set. Performance is measured based on the total number of gradient evaluations required to achieve FSM and TS search convergence. The method (FSM-RIC or FSM-LST) requiring the fewest gradient evaluations is highlighted in bold text.} 
\label{table:birkholz}
\end{table*} 

A test suite consisting of twenty reactions curated by \citet{birkholz2015using} is chosen as the second benchmark. The Birkholz test set contains examples of many types of organic reactions, including insertions, additions, eliminations, hydrolysis, ring-opening, substitutions, cycloadditions, and rearrangements. The description of the twenty reactions in the Birkholz test set is presented in Table \ref{table:birkholz}. This data set shares two identical test cases with the Sharada benchmark set (H$_2$CO $\rightarrow$ H$_2$ + CO and SiH$_2$ + H$_2$ $\rightarrow$ SiH$_4$), resulting in 18 new, unique test cases to evaluate. For each reaction, we highlight in bold the method that locates the TS geometry in fewest calculations.

The comparison between the required gradient calls for FSM and TS calculations using RIC or LST interpolation on the Birkholz test set is presented in Table \ref{table:birkholz}. Both the FSM-RIC and FSM-LST successfully find the exact benchmark TS structure for each test case given the same conservative baseline parameters. The FSM-RIC and FSM-LST again perform similarly, on average, producing a TS guess structure after 58 and 61 gradient evaluations, respectively. The resulting TS guess structures are of similar quality; the FSM-RIC TS guess structure requires on average 21 gradient evaluations for subsequent optimization, and the FSM-LST TS guess structures require on average 24 gradient evaluations for optimization. Of the 18 test cases unique to the Birkholz benchmark set, the FSM-RIC finds the final TS structure in fewer or equal number of gradient evaluations compared to the FSM-LST in 14 of the test cases. 

One notable difference between the performance of the FSM-RIC and FSM-LST on the Birkholz benchmark set occurs in the case of the water-assisted hydrolysis of ethyl acetate to acetic acid and ethanol. Here, an additional water molecule assists in the hydrolysis reaction by proton shuttling which significantly stabilizes the TS structure in comparison to the unassisted reaction mechanism. The TS search for both structures requires a greater number of additional gradient evaluations when compared to other, simpler reactions within the test set. Both the TS guess structure generation and subsequent TS optimization require fewer gradient evaluations when using internal coordinates-based interpolation steps, resulting in a total of 102 and 142 gradient calls required to locate the final TS structure for FSM-RIC and FSM-LST techniques, respectively. 

The Birkholz benchmark data set was originally developed by \citet{birkholz2015using} to test their proposed Connectivity Transition State (CTS) algorithm for determining TS guess structure geometry and TS optimization. In the CTS method, interpolation in redundant internal coordinates limited to bond making and breaking coordinates is performed, and optimization of the interpolated structure is performed on UFF molecular mechanics\cite{casewit1992application} or PM6 semiempirical electronic structure\cite{stewart2007optimization} PESs. Their method also considers different strategies for approximating the initial Hessian used in the subsequent optimization, whereas our TS search calculations are initialized with the exact Hessian calculated at the TS guess structure. This strategy of relying on computationally cheaper levels of theory for optimization results in fewer gradient evaluations overall on the desired PES. This is in contrast to the present work, where the high-level $\omega$B97X-V/def2-TZVP model chemistry is applied in all calculations. While the CTS method is reported to find the final TS structures in significantly fewer high-level gradient evaluations, we report 100\% success rate for both the FSM-RIC and FSM-LST evaluated on the Birkholz benchmark set, whereas the best reported CTS methodology reaches 95\% success rate.

\subsection{Baker test set}

\begin{table*}[ht]
\centering
\resizebox{\textwidth}{!}{\begin{tabular}{|l|l|r|r|r|r|r|r|r|r|}
\toprule
ID & Reaction & \multicolumn{1}{|p{2cm}|}{\centering Gradients \\ (FSM-RIC)} & \multicolumn{1}{|p{2cm}|}{\centering Gradients \\ (TS-RIC)} & \multicolumn{1}{|p{2cm}|}{\centering Gradients \\ (TOTAL-RIC)} & \multicolumn{1}{|p{2cm}|}{\centering Gradients \\ (FSM-LST)} & \multicolumn{1}{|p{2cm}|}{\centering Gradients \\ (TS-LST)} & \multicolumn{1}{|p{2cm}|}{\centering Gradients \\ (TOTAL-LST)}\\
\midrule
1 & HCN $\rightarrow$ HNC & \textbf{78} & \textbf{3} & \textbf{81} & \textbf{72} & \textbf{9} & \textbf{81} \\
2 & HCCH $\rightarrow$ CCH$_2$ & 61 & 11 & 72 & \textbf{63} & \textbf{4} & \textbf{67} \\
3 & H$_2$CO $\rightarrow$ H$_2$ + CO & \textbf{58} & \textbf{13} & \textbf{71} & 62 & 11 & 73 \\
4 & CH$_3$O $\rightarrow$ CH$_2$OH & 52 & 5 & 57 & \textbf{48} & \textbf{7} & \textbf{55} \\
5 & cyclopropyl ring opening & \textbf{59} & \textbf{11} & \textbf{70} & \textit{62} & \textit{25} & \textit{87} \\
6 & bicyclo[1.1.0]butane $\rightarrow$ $\textit{trans}$-butandiene & 61 & 47 & 108 & \textbf{58} & \textbf{25} & \textbf{83} \\
7 & formyloxyethyl 1,2-migration & 63 & 14 & 77 & \textbf{62} & \textbf{11} & \textbf{73} \\
8 & parent Diels-Alder cycloaddition & \textbf{62} & \textbf{15} & \textbf{77} & 53 & 33 & 86 \\
9 & s-tetrazine $\rightarrow$ 2HCN + N$_2$ & \textbf{76} & \textbf{7} & \textbf{83} & 80 & 15 & 95 \\
10 & $\textit{trans}$-butadiene $\rightarrow$ $\textit{cis}$-butadiene & \textbf{63} & \textbf{2} & \textbf{65} & 69 & 2 & 71 \\
11 & CH$_3$CH$_3$ $\rightarrow$ CH$_2$CH$_2$ + H$_2$ & 58 & 24 & 82 & \textbf{52} & \textbf{21} & \textbf{73} \\
12 & CH$_3$CH$_2$F $\rightarrow$ CH$_2$CH$_2$ + HF & \textbf{52} & \textbf{9} & \textbf{61} & 67 & 9 & 76 \\
13 & acetaldehyde keto-enol tautomerism & 63 & 3 & 66 & \textbf{61} & \textbf{4} & \textbf{65} \\
14 & HCOCl $\rightarrow$ HCl + CO & \textbf{65} & \textbf{5} & \textbf{70} & 67 & 6 & 73 \\
15 & H$_2$O + PO$_3^-$ $\rightarrow$ H$_2$PO$_4^-$ & 55 & 21 & 76 & \textbf{48} & \textbf{24} & \textbf{72} \\
16 & CH$_2$CHCH$_2$CH$_2$CHO Claisen rearrangement & \textbf{57} & \textbf{31} & \textbf{88} & 61 & 30 & 91 \\
17 & SiH$_2$ + CH$_3$CH$_3$ $\rightarrow$ SiH$_3$CH$_2$CH$_3$ & \textbf{55} & \textbf{5} & \textbf{60} & 53 & 10 & 63 \\
18 & HNCCS $\rightarrow$ HNC + CS & 72 & 8 & 80 & \textbf{53} & \textbf{9} & \textbf{62} \\
19 & HCONH$_3^+$ $\rightarrow$ NH$_4^+$ + CO & \textbf{65} & 1\textbf{8} & \textbf{83} & 99 & 17 & 116 \\
20 & acrolein rotational TS & \textbf{63} & \textbf{2} & \textbf{65} & 59 & 13 & 72 \\
21 & HCONHOH $\rightarrow$ HCOHNHO & 57 & 12 & 69 & \textbf{62} & \textbf{5} & \textbf{67} \\
22 & HNC + H$_2$ $\rightarrow$ H$_2$CNH & \textbf{36} & \textbf{18} & \textbf{54} & 66 & 15 & 81 \\
23 & H$_2$CNH $\rightarrow$ HCNH$_2$ & \textbf{30} & \textbf{6} & \textbf{36} & \textit{66} & \textit{3} & \textit{69} \\
24 & HCNH$_2$ $\rightarrow$ HCN + H$_2$ & 60 & 41 & 101 & \textbf{53} & \textbf{26} & \textbf{79} \\
\bottomrule
~ & Average & 59 & 14 & 73 & 62 & 14 & 76 \\
\bottomrule
\end{tabular}}
\caption{Comparison of performance of the FSM-RIC and FSM-LST for TS guess structure generation and TS search on the Baker benchmark set. Performance is measured based on the total number of gradient evaluations required to achieve FSM and TS search convergence. The method (FSM-RIC or FSM-LST) requiring the fewest gradient evaluations is highlighted in bold text. Italicized text indicates a failed search or a search where the wrong TS was located.}
\label{table:baker}
\end{table*} 

A test suite consisting of twenty-five reactions originally compiled by \citet{baker1996location} is selected as the third and final benchmark set. The reactions in the test suite again cover a broad set of chemical reaction archetypes, including dissociation, insertion, rearrangement, ring-opening, and rotation reactions. A description of the reactions contained within the test suite, together with TS optimization results, is given in Table \ref{table:baker}. This data set shares eight identical test cases with the Sharada and Birkholz benchmark sets, resulting in 16 new, unique test cases for evaluation. For each reaction, we highlight in bold the method that locates the TS in fewest calculations, and highlight in italics any method resulting in a failed search or a search where the wrong TS was located.

The comparison between the required gradient calls for FSM and TS calculations using RIC or LST interpolation on the Baker test set is presented in Table \ref{table:baker}. The FSM-RIC and FSM-LST perform similarly, on average, producing a TS guess structure after 59 and 62 gradient evaluations, respectively, and both methods requiring on average 14 gradient evaluations for TS optimization. Of the 16 test cases unique to the Baker benchmark set, the FSM-RIC finds the final TS molecular geometry in fewer gradient evaluations compared to the FSM-LST in 12 of the test cases. 

\begin{figure}[t]
\centering
\includegraphics[width=0.8\linewidth]{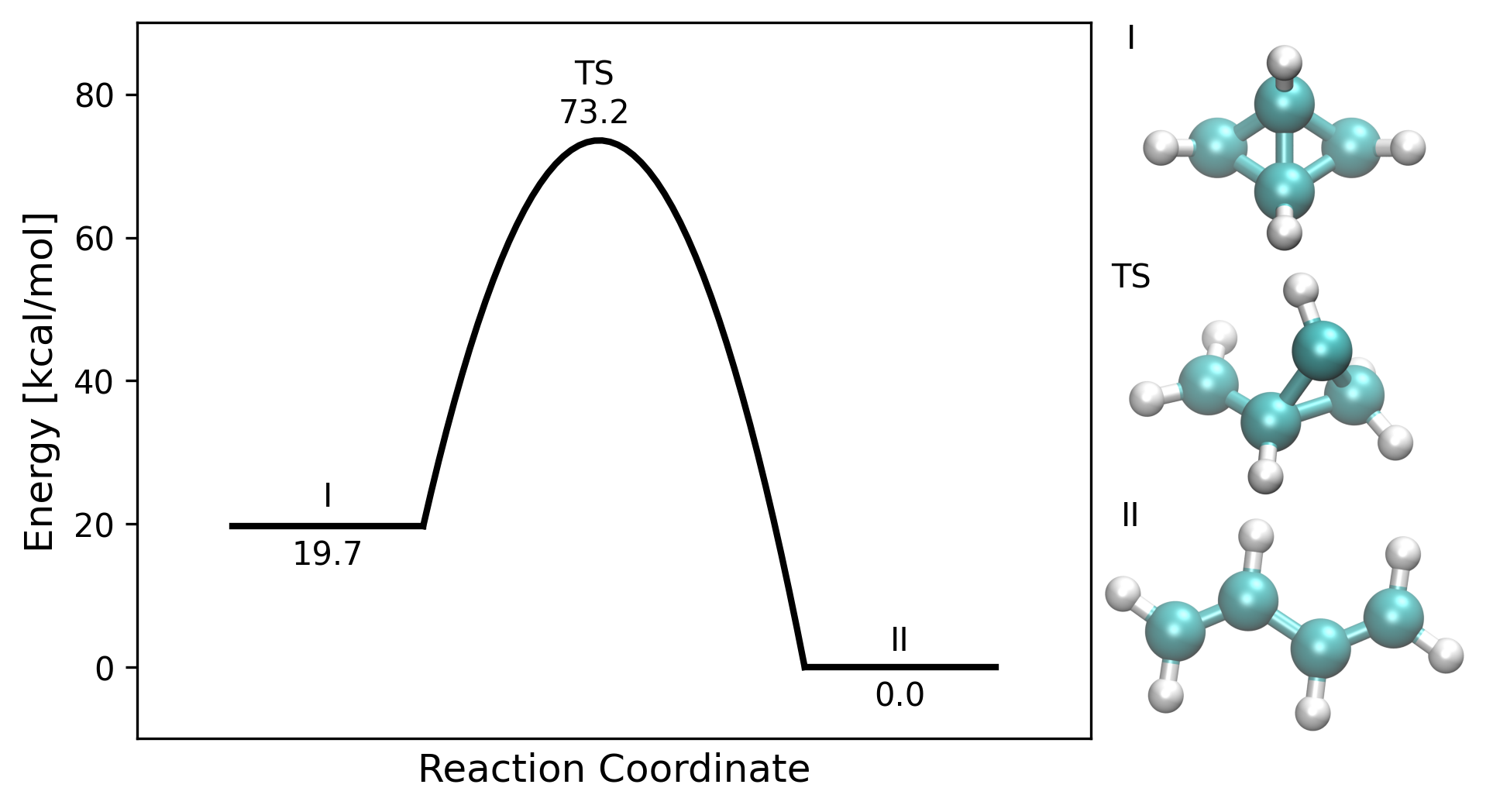}\\
\caption{Reaction coordinate diagram of the bicyclo[1.1.0]butane ring opening reaction to $\emph{trans}$-butadiene.}
\label{fig:bicyclobutane}
\end{figure}

There are two test cases where the TS optimization based on the FSM-RIC requires sizably more gradient evaluations than the equivalent optimization based on FSM-LST, bicyclobutane ring-opening and methylamine dehydrogenation. The reaction coordinate diagram of bicyclobutane ring-opening reaction and structures of the reactant, product, and transition state are shown in Figure \ref{fig:bicyclobutane}. We find the reference transition state to be asymmetric, with one three-membered ring open, and a total barrier height of 73.2 kcal/mol. FSM-LST and FSM-RIC calculations produce TS guess structures with over-estimated barrier heights of 104.8 and 106.1 kcal/mol, respectively. The FSM-LST TS guess structure has two imaginary frequencies with magnitudes of -1035.86 and -179.72 cm${^-1}$, causing the P-RFO algorithm to spend 12 of the 25 optimization steps attempting to correct local Hessian structure and eliminate the second imaginary frequency, while maximizing the energy along the larger magnitude imaginary mode. The FSM-RIC TS guess initially only exhibits one imaginary frequency of -956.75 cm${^-1}$. During optimization of the FSM-RIC TS guess structure, a second imaginary frequency emerges in the updated, approximate Hessian after 8 P-RFO optimization steps. The algorithm then takes 17 additional steps attempting to minimize and eliminate it. The presence of additional imaginary frequencies in both cases accounts for significant computational cost increases during the TS searches, with the LST-based search converging in fewer steps. Similar issues are observed in the methylamine dehydrogenation reaction, where the TS guesses from FSM-LST and FSM-RIC calculations initially have more than one imaginary frequency (FSM-LST) or develop a second imaginary frequency during the P-RFO optimization (FSM-RIC). This results in 14 and 19 additional optimization steps for the LST and RIC-based TS searches respectively. The presence of additional imaginary frequencies in both cases accounts for significant computational cost increases, with the LST-based searches ultimately converging more quickly in these two cases. 

The FSM-RIC successfully finds the exact benchmark TS structure for each test case in the Baker benchmark set. The FSM-LST, however, fails to find the correct TS structure in two of the 16 test cases unique to the Baker benchmark set, as highlighted in italic text in Table \ref{table:baker}. In the case of cyclopropyl radical ring opening reaction, the FSM-RIC TS guess structure is optimized to the correct TS geometry as confirmed by computed TS mode imaginary frequency -1052 cm$^{-1}$ and forward barrier height 53.8 kcal/mol. The FSM-LST TS guess is optimized to a planar C$_3$ molecule with TS mode corresponding to out-of-plane motion of the central carbon atom and associated imaginary frequency -659 cm$^{-1}$ and forward barrier height 103.1 kcal/mol. In the case of methanimine isomerization to aminomethylene (H$_2$CNH $\rightarrow$ HCNH$_2$), the FSM-RIC TS guess structure is optimized to the correct TS geometry where the migrating hydrogen atom moves slightly out-of-plane. The TS geometry is confirmed by computed TS mode imaginary frequency -2210 cm$^{-1}$ and forward barrier height 85.4 kcal/mol. The FSM-LST TS guess structure is optimized in three steps to a second-order saddle point associated with a planar transition geometry. The TS geometry is confirmed to have two vibrational modes with associated imaginary frequencies -2222 cm$^{-1}$ and -472 cm$^{-1}$ and forward barrier height 87.3 kcal/mol. The first vibrational mode corresponds to the correct TS mode while the second vibrational mode is associated with out-of-plane motion.
 
\subsection{Ablation Study}
\begin{figure*}[t]
\centering
\includegraphics[width=\textwidth]{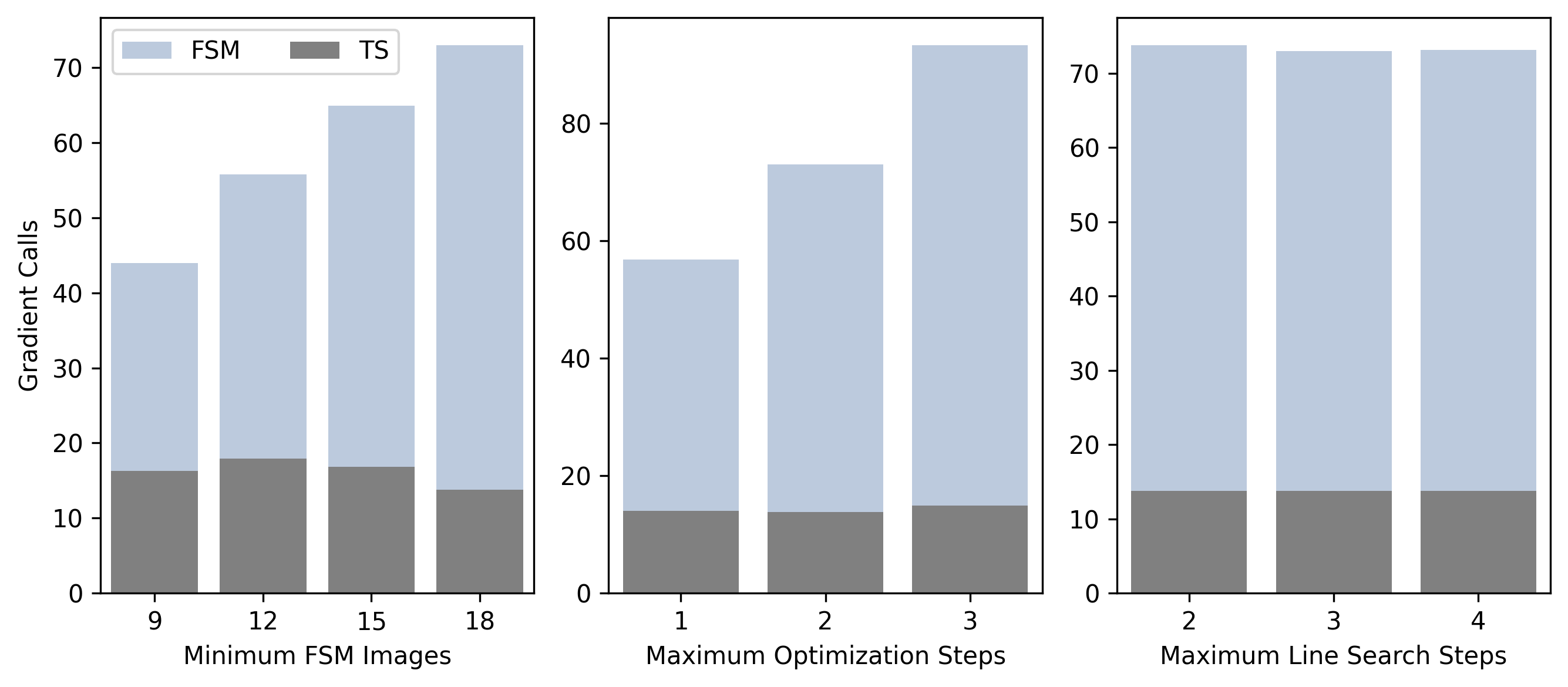}
\caption{
Ablation study of the FSM-RIC for TS guess structure generation and TS search on the Baker benchmark set. Performance is measured based on the average number of gradient evaluations required to achieve TS optimization convergence, and is separated to indicate gradient evaluation used for guess structure generation (FSM) and TS search (TS). (Left) Performance of the FSM-RIC when the minimum number of FSM images (N$_\mathrm{nodes}$) is varied from nine to 18. (Center) Performance of the FSM-RIC when the number of optimization steps per interpolation step (N$_\mathrm{opt}$) is varied from one to three. (Right) Performance of the FSM-RIC when the number of line search steps per optimization step (N$_\mathrm{ls}$) is varied from two to four.
}
\label{fig:ablation}
\end{figure*}

We use the Baker benchmark set for further investigation of the effects of modifying the FSM algorithm parameters on the performance of the FSM-RIC. Figure \ref{fig:ablation} shows the average number of gradient calls required for FSM-RIC calculation and TS optimization per Baker benchmark set test case across three separate ablation studies. In the left panel of Figure \ref{fig:ablation}, we show the results of our first ablation study where the minimum number of string nodes ($N_{\mathrm{nodes}}$) is varied while other optimization parameters are held fixed. We find that decreasing $N_{\mathrm{nodes}}$ from 18 to nine results in a significantly reduced number of gradient calculations for TS guess structure determination, and observe little variation overall in the number of gradient evaluations required for TS optimization. The central and right panels of Figure \ref{fig:ablation} show the results of the ablation studies where $N_{\mathrm{nodes}}$ is held fixed and the maximum number of optimization steps per interpolation step ($N_{\mathrm{opt}}$) and the maximum number of line search steps taken for each optimization step ($N_{\mathrm{ls}}$), respectively, are varied around the original setting. We find that reducing $N_{\mathrm{opt}}$ gives further improvement in the number of required gradient evaluations for TS guess structure determination, while having little effect on the quality of the TS guess structure as indicated by the required gradient evaluations for TS optimization. We find that the required number of gradient evaluations for TS guess structure determination is insensitive to $N_{\mathrm{ls}}$, indicating that though we allow up to three line search steps per optimization step, we often require fewer in practice.

The results of the ablation study suggest that we can significantly reduce the computational cost of TS finding by the FSM-RIC, while maintaining a high success rate, by taking larger interpolation steps (smaller $N_{\mathrm{nodes}}$) and fewer optimization steps per interpolation step. We perform FSM calculations with nominally 9 nodes along the approximate reaction pathway ($N_{\mathrm{nodes}}$), one step per optimization cycle ($N_{\mathrm{opt}}$), and at most three line search steps per optimization cycle ($N_{\mathrm{ls}}$). The comparison between the required gradient calls for FSM and TS calculations using RIC or LST interpolation on the Baker benchmark set with these modified settings is presented in Table \ref{table:baker2}. The FSM-RIC successfully locates the exact benchmark TS structure in fewer required gradient evaluations for each test case. The average required number of gradient evaluations to produce a TS guess structure is 19 and, similar to the more conservative operational settings, the TS guess structures require on average 20 gradient evaluations for optimization. In contrast, the FSM-LST fails to locate the benchmark TS structure in seven of the 24 total test cases present in the Baker benchmark set. The average required number of gradient evaluations to produce a TS guess structure is 19, but TS optimization requires on average 31 gradient evaluations, indicating that the TS guess structures are of worse quality compared with FSM-LST calculations run with more conservative settings. A typical course of action upon TS calculation failure is to re-run the calculation with more conservative settings. This, in practice, would increase the average gradient evaluation requirements for successfully determining a TS structure and can possibly negate the computational cost savings obtained by running the FSM-LST algorithm at more aggressive settings. These results highlight an important finding of this work, that the incorporation of RICs permits the use of larger interpolation steps during TS guess determination, resulting in fewer gradient calculations, while maintaining a good overall success rate when compared with LST interpolation.

\begin{table*}[ht]
\centering
\resizebox{\textwidth}{!}{\begin{tabular}{|l|l|r|r|r|r|r|r|r|r|}
\toprule
ID & Reaction & \multicolumn{1}{|p{2cm}|}{\centering Gradients \\ (FSM-RIC)} & \multicolumn{1}{|p{2cm}|}{\centering Gradients \\ (TS-RIC)} & \multicolumn{1}{|p{2cm}|}{\centering Gradients \\ (TOTAL-RIC)} & \multicolumn{1}{|p{2cm}|}{\centering Gradients \\ (FSM-LST)} & \multicolumn{1}{|p{2cm}|}{\centering Gradients \\ (TS-LST)} & \multicolumn{1}{|p{2cm}|}{\centering Gradients \\ (TOTAL-LST)}\\
\midrule
1 & HCN $\rightarrow$ HNC & \textbf{22} & \textbf{4} & \textbf{26} & 21 & 7 & 28 \\
2 & HCCH $\rightarrow$ CCH$_2$ & \textbf{21} & \textbf{9} & \textbf{30} & 23 & 10 & 33 \\
3 & H$_2$CO $\rightarrow$ H$_2$ + CO & \textbf{18} & \textbf{37} & \textbf{55} & \textit{17} &\textit{53} & \textit{70} \\
4 & CH$_3$O $\rightarrow$ CH$_2$OH & \textbf{16} & \textbf{7} & \textbf{23} & 16 & 8 & 24 \\
5 & cyclopropyl ring opening & \textbf{20} & \textbf{14} & \textbf{34} & \textit{22} & \textit{61} & \textit{83} \\
6 & bicyclo[1.1.0]butane $\rightarrow$ $\textit{trans}$-butandiene & 20 & 74 & 94 & \textbf{20} & \textbf{32} & \textbf{52} \\
7 & formyloxyethyl 1,2-migration & \textbf{24} & \textbf{11} & \textbf{35} & \textit{14} & \textit{62} & \textit{76} \\
8 & parent Diels-Alder cycloaddition & \textbf{23} &\textbf{21} & \textbf{44} & 16 & 30 & 46 \\
9 & s-tetrazine $\rightarrow$ 2HCN + N$_2$ & \textbf{22} & \textbf{8} & \textbf{30} & 25 & 40 & 65 \\
10 & $\textit{trans}$-butadiene $\rightarrow$ $\textit{cis}$-butadiene & \textbf{22} & \textbf{3} & \textbf{25} & \textit{22} & \textit{44} & \textit{66} \\
11 & CH$_3$CH$_3$ $\rightarrow$ CH$_2$CH$_2$ + H$_2$ & 18 & 37 & 55 & \textbf{16} & \textbf{25} & \textbf{41} \\
12 & CH$_3$CH$_2$F $\rightarrow$ CH$_2$CH$_2$ + HF &  \textbf{17} & \textbf{15} & \textbf{32} & 18 & 16 & 34 \\
13 & acetaldehyde keto-enol tautomerism & 20 & 8 & 28 & \textbf{19} & \textbf{8} & \textbf{27} \\
14 & HCOCl $\rightarrow$ HCl + CO & \textbf{22} & \textbf{10} & \textbf{32} & \textbf{20} & \textbf{12} & \textbf{32} \\
15 & H$_2$O + PO$_3^-$ $\rightarrow$ H$_2$PO$_4^-$ & 17 & 34 & 51 & \textbf{16} & \textbf{28} & \textbf{44} \\
16 & CH$_2$CHCH$_2$CH$_2$CHO Claisen rearrangement & \textbf{20} & \textbf{29} & \textbf{49} & 20 & 39 & 59 \\
17 & SiH$_2$ + CH$_3$CH$_3$ $\rightarrow$ SiH$_3$CH$_2$CH$_3$ & 19 & 9 & 28 & \textbf{16} & \textbf{10} & \textbf{26} \\
18 & HNCCS $\rightarrow$ HNC + CS & \textbf{23} & \textbf{8} & \textbf{31} & 26 & 10 & 36 \\
19 & HCONH$_3^+$ $\rightarrow$ NH$_4^+$ + CO & \textbf{20} & \textbf{35} & \textbf{55} & \textit{30} & \textit{1} & \textit{31} \\
20 & acrolein rotational TS & \textbf{24} & \textbf{4} & \textbf{28} & 20 & 17 & 37 \\
21 & HCONHOH $\rightarrow$ HCOHNHO & 19 & 17 & 36 & \textbf{20} & \textbf{11} & \textbf{31} \\
22 & HNC + H$_2$ $\rightarrow$ H$_2$CNH & 6 & 31 & 37 & \textbf{17} & \textbf{17} & \textbf{34} \\
23 & H$_2$CNH $\rightarrow$ HCNH$_2$ & \textbf{9} & \textbf{7} & \textbf{16} & \textit{19} & \textit{5} & \textit{24} \\
24 & HCNH$_2$ $\rightarrow$ HCN + H$_2$ & \textbf{20} & \textbf{46} & \textbf{66} & \textit{19} & \textit{196} & \textit{215} \\
\bottomrule
~ & Average & 19 & 20 & 39 & 20 & 31 & 51 \\
\bottomrule
\end{tabular}}
\caption{Comparison of performance of the FSM-RIC and FSM-LST for TS guess structure generation and TS search on the Baker benchmark set using modified FSM settings (N$_\mathrm{nodes}$=9, N$_\mathrm{opt}$=1, N$_\mathrm{ls}$=3). Performance is measured based on the total number of gradient evaluations required to achieve FSM and TS search convergence. The method (FSM-RIC or FSM-LST) requiring the fewest gradient evaluations is highlighted in bold text. Italicized text indicates a failed search or a search where the wrong TS geometry was located.}
\label{table:baker2}
\end{table*} 

\section{Conclusion}

We have demonstrated the application of the FSM to TS search and optimization using three diverse benchmark sets containing over 40 chemical reactions. The FSM with either LST and RIC interpolation, when run with previously recommended interpolation stepsize and optimization settings, is a reliable technique for producing TS guess geometries and achieves a high success rate for TS optimization on all chemical reactions studied. More importantly, we have shown how incorporation of RIC interpolation into the FSM algorithm, a novel contribution of this work, enables the use of larger interpolation stepsizes and fewer optimization steps per interpolation step, which in turn reduces the computational cost of TS search and optimization while maintaining 100\% success rate. An open-source Python implementation of the FSM algorithm, in addition to the reactant, product, and TS structures of all reactions studied computed at the $\omega$B97X-V/def2-TZVP level of theory is provided as a contribution of this work.\cite{marks2024fsm}

\begin{acknowledgement}
J.G acknowledges start up funding from The University of Iowa. This research was supported in part through computational resources provided by The University of Iowa.
\end{acknowledgement}



\bibliography{fsm}

\end{document}